\documentclass[twocolumn,showpacs,prl,amsmath,amssymb]{revtex4}

\usepackage{graphicx}
\usepackage{dcolumn}
\usepackage{bm}
\usepackage{amsmath,epsfig}

\newcommand{\abs}[1]{\left\vert#1\right\vert}
\newcommand{\ket}[1]{\left\vert#1\right\rangle}
\newcommand{\bra}[1]{\left\langle#1\right\vert}
\newcommand{\braket}[2]{\left\langle#1\vert#2\right\rangle}

\begin{document}


\title{Quantum erasure within the Optical Stern-Gerlach Model}

\author{M. Chianello}
\author{M. Tumminello}
\author{A. Vaglica}%
\author{G. Vetri}%
\affiliation{%
Istituto Nazionale di Fisica della Materia and \\
Dipartimento di Scienze Fisiche ed Astronomiche, Universit\`{a} di Palermo\\
via Archirafi 36, 90123 Palermo, Italy
}%

\date{\today}

\begin{abstract}
In the optical Stern-Gerlach effect the two branches  in which the
incoming atomic packet splits up can display interference pattern
outside the cavity when a field measurement is made which erases
the which-way information on the quantum paths the system can
follow. On the contrary, the mere possibility to acquire this
information causes a decoherence effect which cancels out the
interference pattern. A phase space analysis is also carried out
to investigate on the negativity of the Wigner function and on the
connection between its covariance matrix and the
distinguishability of the quantum paths.
\end{abstract}

\maketitle
\section{\label{sec:level1} I. Introduction}

In a review article of 1991 \cite{Scu} Scully, Englert and Walther
show that ``the information contained in a functioning measuring
apparatus changes the outcome of the experiment''. In the
experiment under consideration, a beam of two-level atoms incident
upon a two-slit arrangement can show interference pattern. Before
to reach the interference region however each atom, initially in
the excited state, passes trough either one of two maser cavities
which are both in the vacuum state, and makes a transition to the
lower state, so that the emitted microwave photon leaves a
which-way information in the cavity. It turns out that the
atom-cavity correlations are responsible for a loss of coherence
which destroys the interference fringes. Following a previous idea
of Scully and Dr\"{u}hl \cite{ScuDru} (also related to the
delayed-choice experiments suggested by Wheeler \cite{w}), they
show how it is possible to retrieve the interference effects by
removing (erasing) the
\textit{welcher Weg} information. \\
The experimental arrangement of the optical Stern-Gerlach  model
\cite{sgo} can be used for a similar quantum optical test of
complementarity. In fact, the interaction Hamiltonian of this
model gives rise to entangled states in which the translational
variables of the atomic center-of-mass are correlated with the
cavity and the internal atomic variables. In this case as well as
in the case analyzed in Ref.\cite{Scu}, the presence or the loss
of coherence in the atomic spatial distribution out the cavity,
can be connected to the peculiar correlations between the
measuring apparatus (the cavity field) and the system being
observed. Diffraction and interference effects induced by field
measurements have been analyzed by Storey, Collett and Walls
\cite{Sto} in the Raman-Nath regime, for virtual atomic
transitions in the presence of very large detuning with respect to
the Rabi frequency. Evoking an old question posed by Popper
\cite{Pop}, namely, if knowledge alone is sufficient to give
uncertainty, they show that the localization of the atom trough
the field measurement may be thought as a creation of a virtual
slit (or slits) and their conclusion is that knowledge alone is
sufficient to create uncertainty. According to this vision, here
we show that the knowledge of the field state erases the which-way
information concerning two mutually exclusive quantum paths the
system can follow. On the contrary, the mere possibility of
acquiring this information which is stored in the
atom-cavity correlations precludes any interference pattern.\\
A phase space analysis is carried out in sec. IV in terms of the
the Wigner distribution for the atomic translational variables.
This analysis allows us to establish a well defined relation
between the determinant of the covariance matrix of the Wigner
function and the distinguishability of the two quantum paths the.\\

\section{\label{sec:level2} II. Model and time evolution of the entire system}

The interaction of a two-level atom with a resonant mode of an
optical cavity is usually described, in the rotating wave
approximation (RWA),  by the well known Jaynes-Cummings
Hamiltonian $\hat{H}_{JC}=\hbar
\omega\,(\hat{a}^{\dag}\hat{a}+\frac{\hat{\sigma}_{z}}{2}+\frac{1}{2})+\hbar
\varepsilon\,(\hat{a}^{\dag}\hat{\sigma}_{-}+\hat{a}\,\hat{\sigma}_{+})
$, where $\hat{\sigma}_{z}$ and $\hat{\sigma}_{\pm}$ account for
the internal atomic dynamics, while $\hat{a}$ and $\hat{a}^{\dag}$
are the annihilation and creation operators for the photons of the
resonant mode of the cavity standing wave, and $\varepsilon$ is
the coupling constant. In this model one assumes that the atom
travels the cavity along a direction which is orthogonal to the
cavity axis, with a velocity sufficiently large to treat
classically this translational degree of freedom. However, even if
this condition is satisfied the translation degree of freedom
along the cavity axis cannot be overlooked. In fact, this dynamics
correlates with the dynamics describing the energy exchange
between the two-level atom and the cavity field, giving rise, for
example, to the optical Stern-Gerlach deflection, to the non
dissipative damping of the Rabi oscillations \cite{Vag1,Cus}, and
so on. Let us therefore consider the optical Stern-Gerlach model
\begin{equation}\label{ham}
  \hat{H}_{SG}=\frac{\hat{p}^{2}}{2\,m}+\hbar \omega \,
  (\hat{a}^{\dag}\hat{a}+\frac{\hat{\sigma}_{z}}{2}+\frac{1}{2})+
  \hbar \varepsilon k \hat{x}\,(\hat{a}^{\dag}\hat{\sigma}_{-}+\hat{a}\,\hat{\sigma}_{+})
\end{equation}
where the dynamics along the x-direction of the cavity axis is
described by the position observable $\hat{x}$ of the atomic
center-of-mass and by its conjugate momentum $\hat{p}$. We have
assumed that the atom, of mass $m$, enters the cavity near a nodal
point of the cavity  k-mode, with a spatial distribution narrow
with respect to wavelength $\lambda=\frac{2 \pi}{k}$ of the
resonant mode ($\Delta x_{0}=\frac{\lambda}{10}$). We also assume
that, at $t=0$, the state of the global system is given by a
factorized form,
\begin{equation}\label{psiinitial}
\ket{\psi(0)}=\ket{\varphi(0)} \ket{e,0}
=\ket{\varphi(0)}\frac{1}{\sqrt{2}}\left(\,\ket{\chi^{+}}+\ket{\chi^{-}}\,\right)
\end{equation}
where the dressed states
$\ket{\chi^{\pm}}=\frac{1}{\sqrt{2}}(\ket{e,0} \pm \ket{g,1})$ are
eigenstates of the excitation number operator,
$(\hat{a}^{\dag}\hat{a}+\frac{\hat{\sigma}_{z}}{2}+\frac{1}{2})\ket{\chi^{\pm}}=\ket{\chi^{\pm}}$
. They are also eigenstates of the interaction energy,
$(\hat{a}^{\dag}\hat{\sigma}_{-}+\hat{a}\,\hat{\sigma}_{+})\ket{\chi^{\pm}}=\pm
\ket{\chi^{\pm}}$ with opposite eigenvalues, while
$\ket{\varphi(0)}$ describes the atomic translational dynamics.
The kets $\ket{e,0}$ and $\ket{g,1}$ denote the atom-cavity states
in which the single excitation of the system pertains to the atom
or to the cavity, respectively. By using the time evolution
operator $\exp(-i H_{SG}\,t/\hbar)$ \cite{Cus} one may write, for
$0\leq t \leq T$ ($T$ is the atomic flight time inside the cavity)
\begin{subequations}\label{psit}
\begin{eqnarray}
\ket{\psi(t)}=\frac{1}{\sqrt{2}}\left(\ket{\phi^{+}(t)}\ket{\chi^{+}}+
\ket{\phi^{-}(t)}\ket{\chi^{-}}\,\right)\label{subeq1} \\
=\frac{1}{2}[\left(\ket{\phi^{+}(t)}+\ket{\phi^{-}(t)}\,\right)\ket{e,0}+\nonumber
\\ \left(\ket{\phi^{+}(t)}-\ket{\phi^{-}(t)}\,\right)\ket{g,1}]\,,\;\;\label{subeq2}
\end{eqnarray}
\end{subequations}
where
\begin{equation}\label{phipmt}
\ket{\phi^{\pm}(t)}=\exp[-\frac{i}{\hbar}(\frac{\hat{p}^{2}}{2\,m}\pm
\hbar \varepsilon\,k
\hat{x})t]\,e^{-i\,\omega\,t}\ket{\varphi(0)}.
\end{equation}
For a Gaussian initial wave packet of minimum uncertainty,
centered in $x_{0}$, with zero mean velocity along the cavity axis
and with a width $\Delta x_{0}$, and getting rid of an irrelevant
global phase factor we have
\begin{eqnarray}\label{phitrapprx}
  \phi^{\pm}(x,t)=\braket{x}{\phi^{\pm}(t)}=[\frac{\Delta x_{0}}{\sqrt{2
  \pi}\beta(t)}]^{\frac{1}{2}}\exp(\mp
  \frac{i}{\hbar} m\,a\,x\,t)\cdot\nonumber \\
  \cdot\exp\{-\frac{(x-x^{\pm}_{t})^{2}}{4\,\beta(t)}\}
\end{eqnarray}
where $a=\frac{\hbar\,\varepsilon\,k}{m}$, $x^{\pm}_{t}=x_{0}\mp
\frac{a}{2}t^{2}$, $\beta(t)=\Delta x_{0}^{2}+\frac{i
\,\hbar}{2\,m}t$, while in the p-representation
\begin{equation}\label{phitrapprp}
  \phi^{\pm}(p,t)=\braket{p}{\phi^{\pm}(t)}=\varphi(p-p^{\pm}_{t},0)
  e^{-i\frac{p\,t}{2\,\hbar}(\frac{p}{m}\mp a\,t)}
\end{equation}
where $p^{\pm}_{t}=\mp m\,a\,t$, and $\varphi(p,0)$ is the Fourier
transform of $\varphi(x,0)$.\\
The Eq.s (\ref{psit}) show that overlap and phase relation between
the two components $\ket{\phi^{\pm}(t)}$ of the atomic
translational state, may play a decisive role in the behavior of
the internal dynamics. On the other hand, they also show that
interference effects in the atomic spatial distribution can appear
in agreement with the record of the photon state in the cavity.\\
For $t \geq T$, while $\phi^{\pm}(p,t)=\phi^{\pm}(p,T)$, the
spatial branches $\phi^{\pm}(x,t)$ evolve according to the free
Hamiltonian, and one has
\begin{widetext}
\begin{equation}\label{phixtT}
  \phi^{\pm}(x,t \geq T)=\left(\frac{\Delta x_{0}}{\sqrt{2\,\pi}\beta(t)}\right)^{\frac{1}{2}}
  \exp\left\{-\frac{[x-\left(x_{0}\mp a\,T(t-T/2)\right)]^{2}}{4\,\beta(t)}
  \mp \frac{i}{\hbar}m\,a\,T\,x\right\}.
\end{equation}
\end{widetext}
As shown by the Eq.s (\ref{psit}) the state of the system splits
up into a coherent superposition of two branches travelling in
opposite directions and encoding correlations between the internal
and the translational dynamics. This suggests to associate the two
orthogonal eigenstates $\ket{\chi^{+}}$ and $\ket{\chi^{-}}$ of
the interaction energy to two mutually exclusive quantum paths,
the two paths actually differing the one from the other for the
opposite direction of the exchanged momentum  between the atom and
the cavity mirrors (Eq. (\ref{phitrapprp})).\\
To clarify this point, let us suppose to perform experiments
measuring the atomic momentum out of the cavity. For
$\varepsilon\,T$ sufficiently large the probability density of
finding a particular momentum $p>0$ ($p<0$) will be different from
zero, as shown by Eq.(\ref{phitrapprp}), only inside a range
$\approx \Delta p_{0}$ around $p_{t}^{-}$ ($p_{t}^{+}$). In fact
for $p>0$ $\abs{\braket{p\,}{\psi(T)}}^{2}\cong
\abs{\phi^{-}(p)}^{2}$ and for $p<0$
$\abs{\braket{p\,}{\psi(T)}}^{2}\cong \abs{\phi^{+}(p)}^{2}$,
while for $p \cong 0$ the probability density is practically zero.
In each case ($p>0$, $p<0$) the measured atomic momentum has been
acquired by the atom from the mirrors of the cavity by means of a
certain number of photon exchanges with the field. For a random
sequence of positive and negative momentum exchanges, the output
distribution of the atomic momentum would be different from zero
in a range $\approx \Delta p_{0}$ around $p=0$. The final
evolution $\ket{\psi(t)}$ for the initial state $\ket{e,0}$ is on
the contrary compatible with a model in which, in a single
experiment where, as above, the atomic momentum is measured, the
atom entering the cavity has first $50$\% probability of
exchanging positive or negative momentum with the mirrors, but
after the first choice, the system keeps to exchange momentum in
the same direction. For either path the atoms exchange momentum
with the mirrors, in such a way as to change their average
momentum from $0$ to $\pm \hbar\,k\,\varepsilon\,T$,
respectively.\\

\section{\label{sec:level3} III. Which way information and quantum erasure}

\begin{figure}
 \includegraphics[width = 0.38 \textwidth]{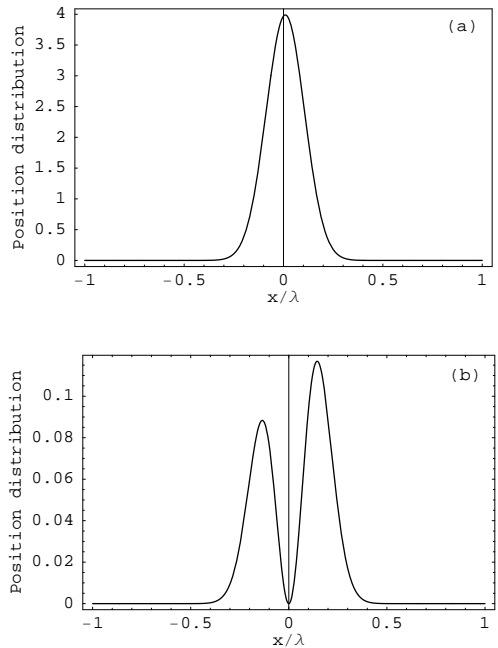}
 \caption{\label{fig2} spatial atomic distributions (\ref{Pmesured0x}) and
(\ref{Pmesured1x}). Fig.s \ref{fig2}a and \ref{fig2}b show that
for $\varepsilon\,T=0.3$ and $t=10\,T$, the probability of finding
the atom in the excited state is enforced by the interference
term, while the complementary probability of finding it in the
ground state is strongly weakened. Concerning the translational
dynamics we suppose an initial wave packet of minimum uncertainty,
with zero mean value of $\hat{p}$, $x_{0}=\lambda/100$ and $\Delta
x_{0}/\lambda=1/10$. The values of the other parameters are
$m=10^{-26}$ kg, $\varepsilon=10^8 sec^{-1}$ and the wavelength
$\lambda=10^{-5}$ meters.}
\end{figure}

\begin{figure}[t]
 \includegraphics[width = 0.38 \textwidth]{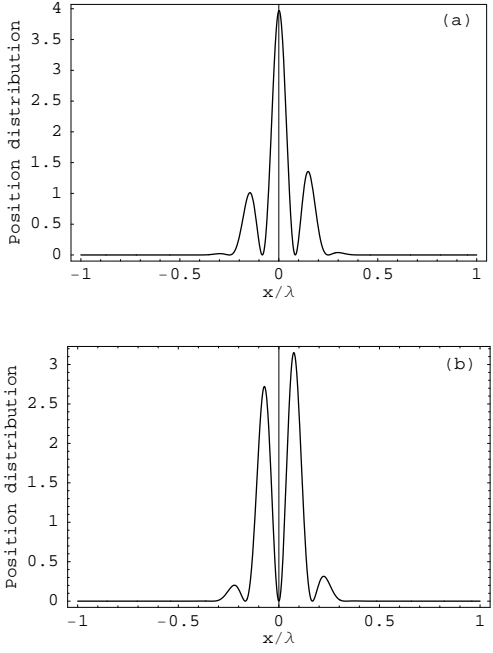}
 \caption{\label{fig3} spatial atomic distributions (\ref{Pmesured0x}) and
(\ref{Pmesured1x}). Fig.s \ref{fig3}a and \ref{fig3}b show that
for $\varepsilon\,T=3$ and $t=10\,T$, the interference term gives
rise to fringes (Fig.\ref{fig3}a) and antifringes
(Fig.\ref{fig3}b) patterns. The values of the other parameters are
the same of Fig.\ref{fig2}}
\end{figure}

\begin{figure}[t]
 \includegraphics[width = 0.38 \textwidth]{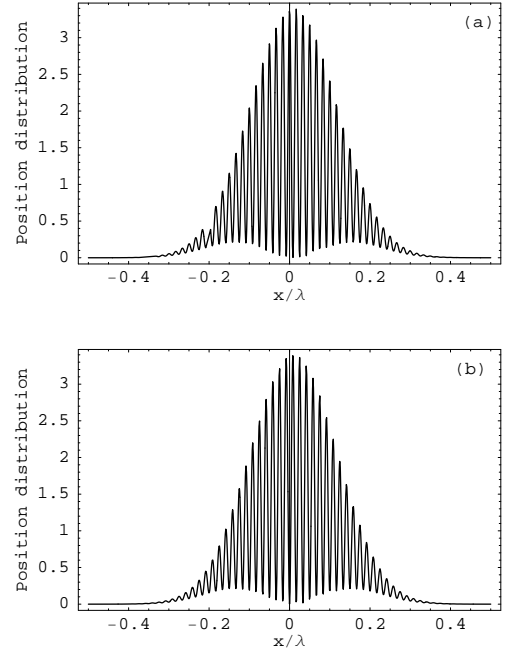}
 \caption{\label{fig4} spatial atomic distributions (\ref{Pmesured0x}) and
(\ref{Pmesured1x}). Fig.s \ref{fig4}a and \ref{fig4}b show that
for $\varepsilon\,T=30$ and $t=10\,T$, the oscillations in $x$ of
the interference term occur in closer succession. The values of
the other parameters are the same of Fig.\ref{fig2}}
\end{figure}

As in the double-slit experiment of Young, we may observe
interference fringes if there is no way to acquire information on
the quantum paths. In our case this information erasure can be
accomplished, for example, by measuring the cavity photon state.
In fact, the states $\ket{e,0}$, and $\ket{g,1}$, are associated
to the atom going through both the quantum paths, with definite
relative phase, different for the two states. \\
On the contrary, if the above measurement is not made, from
Eq.(\ref{subeq1}), using the orthonormality of the dressed states,
we get a spatial atomic distribution,  both inside and out of the
cavity,
\begin{equation}\label{Px}
  P(x)=\abs{\braket{\psi(t)}{x}}^{2}=\frac{1}{2}\left(\abs{\phi^{+}(x,t)}^{2}+
  \abs{\phi^{-}(x,t)}^{2}\right),
\end{equation}
which does not exhibit interference terms since the correlations
between the cavity and the atom cause a loss of coherence which
destroys the interference.
One can also say that because of the entanglement, the atom
imprints in the cavity information about the followed path, which
is encoded in the interaction energy and that could also be read
by finding out the direction of the exchanged momentum with the
mirrors. This information serves as a which-way identification and
leads, as will be shown later, to a classical-like probability
distribution in the phase space. In this case, it is the mere
possibility of recovering the information on the direction side of
the exchanged
momentum that makes distinction between the two paths.\\
Let us now suppose that a measurement of the photon field in the
cavity is made which records zero photons in the cavity, or one
photon. The atomic state out of the cavity, after the measurement,
will be proportional, respectively, to
\begin{subequations}\label{psiMesuret}
\begin{eqnarray}
  \ket{\psi_{0}(t)}= \frac{1}{2}\left(\ket{\phi^{+}(t)}+\ket{\phi^{-}(t)}\right)\ket{e},\label{psi0t}\\
  \ket{\psi_{1}(t)}= \frac{1}{2}\left(\ket{\phi^{+}(t)}-\ket{\phi^{-}(t)}\right)\ket{g}.\label{psi1t}
\end{eqnarray}
\end{subequations}
 \\
If we repeat the measurement for many atoms entering the cavity in
the same conditions, the interference patterns of the atoms
correlated to the two different records will consequently be given
respectively by
\begin{widetext}
\begin{subequations}\label{Pmesuredx}
\begin{eqnarray}
  P_{0}(x)= \frac{1}{4}\left(\abs{\phi^{+}(x,t)}^{2}+\abs{\phi^{-}(x,t)}^{2}+
                      \phi^{+}(x,t)^{*}\phi^{-}(x,t)+ c.c.\right),\label{Pmesured0x}\\
  P_{1}(x)=  \frac{1}{4}\left(\abs{\phi^{+}(x,t)}^{2}+\abs{\phi^{-}(x,t)}^{2}-
                      \phi^{+}(x,t)^{*}\phi^{-}(x,t)- c.c.\right),\label{Pmesured1x}
\end{eqnarray}
\end{subequations}
\end{widetext}
where $\phi^{\pm}(x,t)$ are given by the Eq. (\ref{phixtT}).\\
Eq.s (\ref{psiMesuret}) and (\ref{Pmesuredx}) show that the
measurement of the photon field in the cavity plays the role of
``quantum eraser" which removes the which way information from the
cavity and retrieves the interference terms. It is worth to
observe that for the two different records different interference
effects will appear in
form of ``fringes" and ``antifringes" patterns.\\
It is however to point out that the distinguishability of the two
quantum paths depends on the momentum distribution as induced by
the field interaction. In particular, the measurement of a
positive or negative momentum exchange with the cavity could be
possible only if the distance between the two main peaks is
distinctly larger than the uncertainty $\Delta p_{0}$, that is
$\hbar\,k\,\varepsilon\,T > \Delta p_{0}$. Since in our case we
assume $\Delta x_{0}=\lambda/10$, we have $\hbar\,k \approx
1.26\,\Delta p_{0}$ and the two paths can be distinguished if
$\varepsilon\,T > 1$. When this condition is not satisfied, the
impossibility of distinguishing the two paths gives rise, out of
the cavity, to spatial atomic distribution which exhibit
diffraction rather than interference patterns (See Fig.
\ref{fig2}).\\
To analyze the spatial atomic distribution given by the Eq.s
(\ref{Pmesuredx}), it is convenient to examine first the
interference term
\begin{widetext}
\begin{equation}\label{2RePhiupPhidown}
2\,Re
\left[\phi^{+}(x,t)^{*}\phi^{-}(x,t)\right]=\frac{2}{\sqrt{2\,\pi}\Delta
x_{t}}\exp\left\{-\frac{(x-x_{0})^{2}}{2\,\Delta x_{t}^{2}}-\frac{
\left[a\,T (t-T/2) \right]^{2}}{2\,\Delta x_{t}^{2}}
\right\}\cos\{2\,\varepsilon\,T\,k[x-(x-x_{0})\eta(T,t)]\}
\end{equation}
where
\begin{equation}\label{parameters}
  \eta(T,t)=\frac{\Delta p_{0}^{2}t\,(t-T/2)}{m^{2}\Delta
  x_{t}^{2}},\qquad \Delta x_{t}^{2}=\Delta x_{0}^{2}+\frac{\Delta
  p_{0}^{2}}{m^{2}}t^{2},\qquad \Delta p_{0}=\frac{\hbar}{2\,\Delta
  x_{0}}.
\end{equation}
\end{widetext}
The Eq. (\ref{2RePhiupPhidown}) shows that for short flight time
out of the cavity ($t\geq T$) and for used values of the
parameters, the interference term is of the same order of the main
peaks $\abs{\phi^{\pm}(x,t)}^{2}$ of the spatial distribution, and
fringe patterns can be observed. For increasing free flight time
$t \gg T$ and $\varepsilon\,T >1$, the main peaks travel far away
the one from the other, while the interference term is damped to
zero and the fringe patterns disappear. In fact for $t \gg T$, the
exponential factor of Eq. (\ref{2RePhiupPhidown}) gives rise to a
damping for
\begin{equation}\label{condDemp}
\frac{1}{2}\left[\left(\frac{\Delta x_{0}}{a\,t\,T}\right)^{2}+
\left(\frac{\Delta p_{0}}{m\,a\,T}\right)^{2}\right]^{-1} > 1
\end{equation}
This implies that both the terms in the bracket be smaller than
$1$, that is, the main peaks must be well separated in both the
position and momentum space. The condition on the momentum space
does not depend on $t$ and implies $\varepsilon\,T>1$ If this last
condition is satisfied, the damping takes place for $\Delta
p_{0}\,t/m \cong \Delta x_{0}$ and consequently, for the
parameters used in Fig.s \ref{fig3},
$\pi\,\varepsilon\,T\,t>10^{-3}$. The loss of the interference
fringes is in these cases only due to the increasing distance
between the two paths whose superposition is still coherent. As a
consequence, as will be shown in the next section, the probability
distribution in the phase space contains oscillating terms which
show disagreement with a classical like behavior.\\
The spatial atomic distributions (\ref{Pmesured0x}) and
(\ref{Pmesured1x}) are visualized in Fig.s \ref{fig2}a,
\ref{fig3}a, \ref{fig4}a and \ref{fig2}b, \ref{fig3}b,
\ref{fig4}b, respectively, as a function of $x/\lambda$, for
different values of $\varepsilon\,T$. Fig.s \ref{fig2} show that
for $\varepsilon\,T=0.3$ and $t=10\,T$, the probability of finding
the atom in the excited state is enforced by the interference
term, while the complementary probability of finding it in the
ground state is strongly weakened.
The spatial distributions for $t \gg T$
do not change sensitively.\\
For $\varepsilon\,T=3$ and $t=10\,T$,  the interference term gives
rise to fringes (Fig.\ref{fig3}a) and antifringes
(Fig.\ref{fig3}b) patterns which are evident up to $t\cong
5\,10^{3} T$, while for much longer times disappear.\\ For
$\varepsilon\,T=30$, the oscillations in $x$ of the interference
term occur in closer succession and the
resulting patterns are shown in Fig.s \ref{fig4}a and \ref{fig4}b.\\
\section{\label{sec:level4} IV. Phase Space analysis: determinant of the covariance matrix}

Some properties of the atomic spatial distribution and of its
correlation to the internal dynamics can be better understood by
means of the Wigner function $W(x,p)$ \cite{Hill}.\\
We first analyze the Wigner function of the reduced density
operator
\begin{eqnarray}\label{rhorid}
  \hat{\rho}_{r}=Tr_{i}\ket{\psi(t)}\bra{\psi(t)}=\qquad\qquad\qquad\qquad\nonumber\\
  =\frac{1}{2}\left[\ket{\phi^{+}(t)}\bra{\phi^{+}(t)}+
  \ket{\phi^{-}(t)}\bra{\phi^{-}(t)}\right],
\end{eqnarray}
where $\ket{\psi(t)}$ is given by Eq. (\ref{subeq1}) and $Tr_{i}$
indicates the trace on the internal atomic-cavity states. The
orthogonality of the dressed states leads to a reduced operator
which appear as the incoherent sum of two terms each one related
to a particular component of the translational state. Using the
characteristic function associated to $\hat{\rho}_{r}$,
\begin{equation}\label{characteristic}
  C(\lambda_{x},\lambda_{p})=Tr \left \{\hat{\rho}_{r} \exp\left[\frac{i}{\hbar}
  (\lambda_{x}\hat{x}+\lambda_{p}\hat{p})\right]\right\}
\end{equation}
and $\phi^{\pm}(x,t)$ as given by Eq. (\ref{phixtT}), we derive
the Wigner function $W_{r}(x,p)$ as Fourier transform of
$C(\lambda_{x},\lambda_{p})$ and we obtain (for $t > T$)
\begin{equation}\label{WignerR}
  W_{r}(x,p)=W^{+}(x,p)+W^{-}(x,p),
\end{equation}
where
\begin{eqnarray}\label{WignUpDown}
W^{\pm}(x,p)=\frac{1}{2\pi\hbar}\exp\left\{-\frac{\left[\frac{[x\pm
a T(t-T/2)]^{2}}{\Delta x_{t}^{2}}+\frac{(p\pm m a T)^{2}}{\Delta
p_{0}^{2}}\right]}{2(1-\rho_{xp}^{2})}\right\}\cdot\nonumber \\
\cdot\exp\left\{\frac{\rho_{xp}\frac{(p\pm m\,a\,T)}{\Delta
p_{0}}\frac{[x\pm a\,T(t-T/2)]}{\Delta
x_{t}}}{2(1-\rho_{xp}^{2})}\right\},\;\;\nonumber
\end{eqnarray}
$\rho_{x p}$ is the free particle correlation coefficient of the
atomic translational variables
\begin{eqnarray}\label{CoeffCorr}
\rho_{x p}=\frac{cov(x,p\,;t)}{\Delta x_{t}\Delta
p_{0}}=\frac{\Delta p_{0}\,t}{m\,\Delta x_{t}},\nonumber \\
cov(x,p\,;t)=\frac{1}{2}\left\langle\hat{x}\hat{p}+\hat{p}\,\hat{x}\right\rangle_{t}-
\left\langle\hat{x}\right\rangle_{t}\left\langle\hat{p}\right\rangle_{t}.
\end{eqnarray}
\\
and we have put $x_{0}=0$.\\
As in Eq.(\ref{Px}), any phase relation between $\phi^{+}(x,t)$
and $\phi^{-}(x,t)$ has been lost after the trace operation on the
internal dynamics, and as a consequence the Wigner function
assumes only positive values and can figure a classical-like
joined probability distribution in the phase space.\\
The classicality of the probability distribution has been here
recovered by releasing information on the internal dynamics and
could be considered as a natural ending up of the distribution if
a decay time of the photon in the cavity would be taken into
account, after which, the information on the cavity state is
irreversibly lost.\\
The positive definition of the Wigner function allows to comment
on the generalized uncertainty area $\Delta$ (the error box
\cite{Cav}) of the whole spatial distribution in the phase space
and to relate it to the distance between the gaussian
distributions which describe the different paths. The square
uncertainty $\Delta^{2}$ is evaluated as the determinant of the
distribution covariance matrix
\begin{equation}\label{errboxGen}
\Delta(t)^{2}=\Delta x(t)^{2}\Delta p(t)^{2}-cov(x,p\,;t)^{2}
\end{equation}
and for $t < T$ it is given by
\begin{equation}\label{DeltaD}
  \Delta(t)^{2}=\frac{\hbar^{2}}{4}\left(1+\frac{D(t)^{2}}{4}\right),
\end{equation}
where
\begin{equation}\label{Dt}
D(t\leq
T)=2\,\sqrt{\left(\frac{a\,t^{2}}{2}\right)^{2}\frac{1}{\Delta
x_{0}^{2}}+\frac{(m\,a\,t)^{2}}{\Delta p_{0}^{2}}}
\end{equation}
is the adimensional distance in the phase space between the
average positions of the two gaussian centered in
$(x^{+}_{t},p^{+}_{t})$ and $(x^{-}_{t},p^{-}_{t})$, measured in
units of $\Delta x_{0}$ and $\Delta p_{0}$ along $x$ and $p$,
respectively. It is to notice that $D(t \ge T)=D(T)$ and then
$\Delta(t > T)=\Delta(T)$. In fact, for $t > T$ the atom evolves
freely and the uncertainty area (or the determinant of the
covariance matrix) is constant in time as it happens for particle
subjected to a potential at most quadratic. Since $\hbar^{2}/4$ is
the square of the initial uncertainty area, the enlargement of the
square area in time is only due to the distance $D(t)$.\\
The distance $D(t)$ in the phase space plays a fundamental role in
determining the distinguishability of the two quantum paths
related to the two translational branches $\ket{\phi^{+}(t)}$ and
$\ket{\phi^{-}(t)}$. In fact, the following relation holds
\begin{equation}\label{scalar}
  \braket{\phi^{+}(t)}{\phi^{-}(t)}\propto
  \exp\left(-\frac{D(t)^{2}}{8}\right),
\end{equation}
which shows how the two atomic translational paths become mutually
exclusive for increasing values of $D$.\\
We want now investigate the effect of a measurement of the photon
field in the cavity on the Wigner function. The translational
system state after the measurement will be proportional to one of
the states (\ref{psiMesuret}):
\begin{subequations}\label{psiAfterMes}
\begin{eqnarray}
  \braket{x}{\psi^{0}(x,t)}=\frac{1}{\sqrt{N_{0}(T)}}\left[\phi^{+}(x,t)+\phi^{-}(x,t)\right]\label{psiAM0}\\
  \braket{x}{\psi^{1}(x,t)}=\frac{1}{\sqrt{N_{1}(T)}}\left[\phi^{+}(x,t)-\phi^{-}(x,t)\right],\label{psiAM1}
\end{eqnarray}
\end{subequations}
where
\begin{subequations}\label{Normalization}
\begin{eqnarray}
N_{0}(T)=2\left[1+e^{-D(T)^{2}/8}\cos(2 \varepsilon k x_{0} T) \right],\label{N0}\\
N_{1}(T)=2\left[1-e^{-D(T)^{2}/8}\cos(2 \varepsilon k x_{0} T)
\right],\label{N1}
\end{eqnarray}
\end{subequations}
(For seek of simplicity, we will assume again that $x_{0}=0$). The
Eq.s (\ref{psiMesuret}) show that after the measurement of the
photon field, the spatial atomic state is given by a linear
superposition of two gaussians, which, as well known \cite{Ball},
gives rise to negative contributions to the quasi probability
distribution. By using in this case a more commune definition of
the Wigner function,
\begin{equation}\label{WignerStandForm}
  W(x,p)=\frac{1}{2\,\pi\,\hbar}\mathop{\int}_{-\infty}^{+\infty} \psi(x+\frac{\delta}{2},t)^{*}
  \psi(x-\frac{\delta}{2},t)e^{-i
  \frac{\lambda\,p}{\hbar}}d\delta,
\end{equation}
one easily gets for the two states (\ref{psiAfterMes})
\begin{subequations}
\begin{eqnarray}
W_{0}(x,p)=\frac{2}{N_{0}(T)}\left[W^{+}(x,p)+W^{-}(x,p)+W^{q}(x,p)\right],\:\,\label{W0am}\\
W_{1}(x,p)=\frac{2}{N_{1}(T)}\left[W^{+}(x,p)+W^{-}(x,p)-W^{q}(x,p)\right],\:\,\label{W1am}
\end{eqnarray}
\end{subequations}
where
\begin{widetext}
\begin{equation}\label{Wq}
W^{q}(x,p)=\frac{1}{2\,\pi\,\hbar}\exp\left[-\frac{1}{2(1-\rho_{xp}^{2})}
\left(\frac{x^{2}}{\Delta x_{t}^{2}}+\frac{p^{2}}{\Delta
p_{0}^{2}}+2\,\rho_{x p}\frac{p}{\Delta p_{0}}\frac{x}{\Delta
x_{t}}\right)\right]\,2\,\cos\left\{2\,\varepsilon\,k\,T\left[x+\frac{p}{m}
\left(t-\frac{T}{2}\right)\right]\right\},
\end{equation}
\end{widetext}
gives negative contributions and testifies a precise phase
relation between the two components $\phi^{\pm}$, and the
consequent possibility of interference fringes in the spatial
atomic distribution. $W^{q}(x,p)$ is of the same order of
magnitude of the two main peaks which travel in opposite direction
in the phase space, even when they are absolutely disconnected,
that is even in a ``macroscopical limit'' in which the scalar
product given by the Eq. (\ref{scalar}) is practically zero
($\varepsilon\,T \gg 1$).\\
Supposing $0$ photons in the cavity or $1$ photon as the output of
the measure, the following expressions for the square of the
generalized uncertainty area is obtained:
\begin{widetext}
\begin{equation}\label{DeltaMesured}
  \Delta^{\binom{0}{1}}(t \geq
  T)^{2}=\frac{\hbar^{2}}{N_{0 \atop 1}(T)^{2}}\left\{1+\frac{D(T)^{2}}{4}\pm
  \left[2-\left(\frac{D(T)^{2}}{4}\right)^{2}\right]\,e^{-D(T)^{2}/8}+\left(1-\frac{D(T)^{2}}{4}\right)\,e^{-D(T)^{2}/4}
\right\}
\end{equation}
\end{widetext}
The error box of the ``classical like'' distribution given in the
Eq. (\ref{DeltaD}) is smaller or larger than $\Delta^{(1)}(t \geq
T)^{2}$ or $\Delta^{(0)}(t \geq T)^{2}$, respectively. However,
for $\varepsilon T \gg 1$, both $\Delta^{(0)}(t \geq  T)$ and
$\Delta^{(1)}(t \geq T)$ tend to the same value of the uncertainty
area related to the ``classical
like'' distribution of Eq. (\ref{WignerR}).\\
\section{\label{sec:level5} V. Conclusions}

It has been shown that in the optical Stern-Gerlach effect, if the
information on the internal dynamics is released, the atom leaves
the cavity in a state which results an incoherent superposition of
two spatial paths related to two virtual slits. However, the
measurement of photon presence in the cavity erases the which path
information and spatial atomic interference may occur from the
virtual slits, which exhibit fringe and antifringe patterns in
accord to the lack or the presence of the photon. The Wigner
quasi-probability atomic distribution in the phase space, exhibits
in the first case a ``classical like'' behavior, characterized by
a superposition of two incoherent distributions related to the two
paths. The Wigner function after the quantum erasure shows on the
contrary the positive-negative oscillations typical of a coherent
superposition of two quantum paths. Quite intriguing is however
the fact that the error box of this second Wigner distribution
tends to the ``classical like'' one when the adimensional distance
in the phase space between the two paths is much larger than $1$
(one could say in a sort of macroscopic limit). In fact, when the
oscillation lengths in the momentum space (or in the co-ordinate
space)  of the Wigner distribution are much smaller than the
initial uncertainties of the atomic packet, the oscillating part
does not contribute to the evaluation of the second moments of
both $x$ and $p$.

\end{document}